\documentclass[aps, pra, twocolumn, superscriptaddress, amsmath, tightenlines, longbibliography]{revtex4-2}

\usepackage{amssymb}
\usepackage{amsmath}
\usepackage{cases}
\usepackage{dcolumn}
\usepackage{graphicx}
\usepackage{mathrsfs}
\usepackage{appendix}
\usepackage{graphicx}
\usepackage{booktabs}
\usepackage{colortbl}
\usepackage{float}
\usepackage{ulem}

\definecolor{Dred}{RGB}{190,0,0}

\usepackage{url}
\usepackage[colorlinks]{hyperref}
\hypersetup{%
	plainpages=true,
	breaklinks=true,       
	hypertexnames=false,  
	pageanchor=true,
	colorlinks=true,
	linkcolor={blue},
	citecolor={red},
	urlcolor={blue},
	anchorcolor={black}
}

\def \hide#1{}

\begin{document}
	\title{Amplifying Decoherence-Free Many-Body Interactions with Giant Atoms Coupled to Parametric Waveguide}
	
	\author{Xin Wang}
	\affiliation{School of Physics, Xi'an Jiaotong University, Xi'an 710049, 
		People’s Republic of China}
	
	\author{Zhao-Min Gao}
	\affiliation{School of Physics, Xi'an Jiaotong University, Xi'an 710049, 
		People’s Republic of China}

	\date{\today}
	
	\begin{abstract} 
		Parametric amplification offers a powerful means to enhance quantum interactions through field squeezing, yet it typically introduces additional noise which accelerates quantum decoherence, a major obstacle for scalable quantum information processing. The squeezing field is implemented in cavities rather than continuous waveguides, thereby limiting its scalability for applications in quantum simulation. Giant atoms, which couple to waveguides at multiple points, provide a promising route to mitigate dissipation via engineered interference, enabling decoherence-free interactions. We extend the squeezing-amplified interaction to a novel quantum platform combining giant atoms with traveling-wave parametric waveguides based on $\chi^{(2)}$ nonlinearity. By exploiting destructive interference between different coupling points, the interaction between giant atoms is not only significantly enhanced but also becomes immune to squeezed noise. Unlike conventional waveguide quantum electrodynamics without a squeezing pump, the giant emitters exhibit both exchange and pairing interactions, making this platform particularly suitable for simulating many-body quantum physics. More intriguingly, the strengths of these interactions can be smoothly tuned by adjusting the squeezing and coupling parameters. Our architecture thus provides a versatile and scalable platform for quantum simulation of strongly correlated physics and paves the way toward robust quantum control in many-body regimes.
		
	\end{abstract}
	\maketitle
	\section{Introduction} Achieving strong interactions across diverse platforms is a central goal in quantum optics, as it enables a wide range of applications, including quantum state manipulation, simulation of many-body physics, and high-precision quantum sensing~\cite{PhysRevB.70.205112,Frisk_2019,PhysRevLett.95.036401,PhysRevB.94.155418,PhysRevResearch.3.023093,PhysRevLett.96.057405}. Quantum amplification using squeezed light provides a promising route to strong or even ultrastrong coupling~\cite{PhysRevLett.126.023602,PhysRevLett.133.033603,PhysRevA.107.053711,PhysRevA.108.032411,PhysRevA.105.062425,PRXQuantum.5.020314,PRXQuantum.5.020306,nature7160,science.aaw2884,Qiu_2023,PhysRevLett.114.093602}. However, the squeezing process unavoidably introduces enhanced parametric noise, which leads to rapid decoherence~\cite{Qin2024,RevModPhys.84.1,PhysRevLett.118.103601,PhysRevA.102.063715}. To mitigate this issue, additional experimental techniques are often required, such as injecting a carefully designed squeezed vacuum field into the waveguide or performing frequent measurements~\cite{Xia2024,PhysRevLett.120.093602,Squeezing,PhysRevA.100.012339,PhysRevLett.120.093601,PhysRevA.98.033818}, which introduce additional experimental overheads. Moreover, previous studies on squeezing amplification have been confined to cavity quantum electrodynamics (cavity-QED) and few-body interactions, rather than exploring many-body physics in quantum networks~\cite{PhysRevA.110.052433,cai2025}. These limitations highlight the urgent need for a novel squeezing amplification paradigm that is free from parametric noise as well as capable of long-distance strong interactions.
	
	Giant atoms, characterized by nonlocal multipoint coupling, have emerged 
	as a promising platform in waveguide QED for addressing these 
	challenges~\cite{Nature_superconducting,PhysRevA.103.023710,Wang_squid,Wang_2022,JQYou_2022}.
	 Unlike conventional atoms whose coupling is point-like and inevitably 
	leads to radiative loss, giant atoms can exhibit decoherence-free effect 
	across their entire Hilbert space due to destructive interference between 
	spatially separated coupling points, while still maintaining coherent 
	interactions~\cite{PhysRevA.108.013704,PhysRevA.101.053855,PhysRevResearch.6.043222,PhysRevLett.120.140404,PhysRevResearch.2.043184,PhysRevA.107.023705}.
	 This unique feature enables scalable, long-range interactions without 
	the need for cavities, and the interference effects may further be 
	harnessed to suppress parametric noise while preserving coherent dynamics.
	
	Therefore, in this work, we propose a platform that combines traveling-wave parametric waveguide QED with giant atoms. $\chi^{(2)}$-based traveling-wave parametric amplifier waveguides use second-order optical nonlinearity to efficiently amplify light, convert frequencies and generate quantum states like squeezed light without resonators~\cite{Anderson_97,Eto_08,Rizvanov_2024}. By harnessing the intrinsic destructive interference of giant atoms, our proposal enables coherent interactions free of squeezing noise beyond single-cavity confinement. Crucially, it supports tunable exchange and pairing interactions, both of which can be significantly enhanced by large parametric gain and modulated via the relative phase of pump fields. In contrast to single-mode cavity QED, where strong coupling typically yields fixed collective interactions, our waveguide platform enables tunable, spatially structured many-body interactions, providing a scalable and noise-resilient route to programmable quantum simulation of strongly interacting quantum phenomena.
	
	\begin{figure*}
		\centering
		\includegraphics[width=0.95\linewidth]{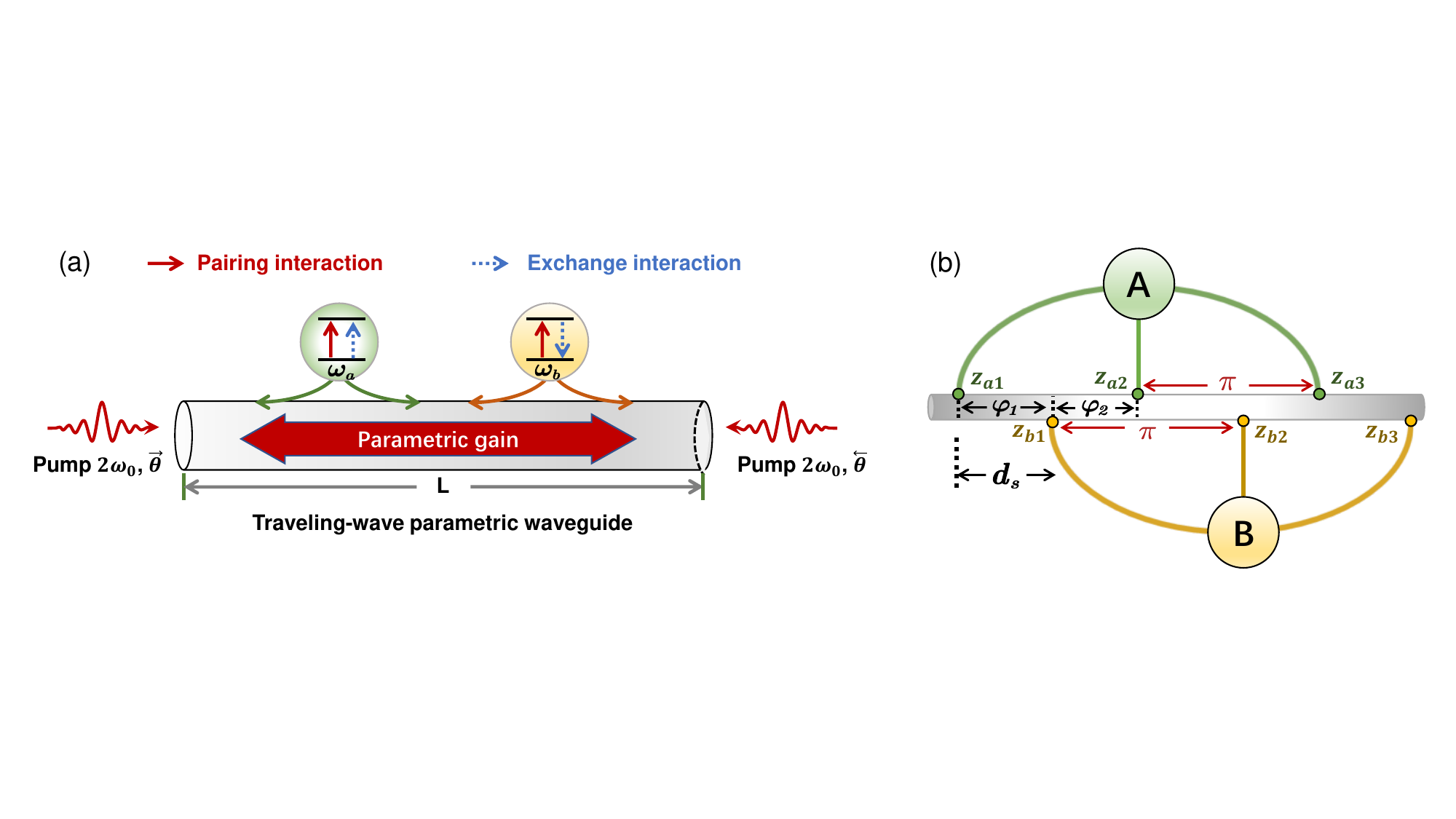}
		\caption{(a) Scheme of two giant atoms coupled to a nonlinear traveling-wave parametric waveguide of length $L$, pumped by counter-propagating fields at $2\omega_0$ to generate bidirectional squeezed vacuum fields, with pump phases $\overset{\rightarrow}{\theta}$ and $\overset{\leftarrow}{\theta}$. Blue dotted arrow denotes the coherent-exchange interaction $J_c$. The red dashed arrow represents the pairing interaction $J_p$. (b) Scheme of a pair of giant atoms coupled to the waveguide at three equidistant points by distance $\pi$. And the phase difference is $k_0\varphi_1$, with $\varphi_1+\varphi_2=\pi$.}
		\label{fig1}
	\end{figure*}
	
	\section{Model}
	\subsection{Giant atoms interacting with nonlinear parametric waveguide}
	In conventional waveguide QED, waveguides are typically treated as vacuum or thermal channels, leading to dissipative and incoherent interactions between atoms coupled to the waveguide, without parametric process. However, in many-body simulations in condensed matter physics, pairing interactions are often essential. To address this, we consider a nonlinear parametric waveguide, which is possible to mediate such coherent pairing interactions through engineered parametric process.
	
	We first consider a minimal setup consisting of two giant atoms with frequencies $\omega_a$ and $\omega_b$ coupled to a parametric waveguide of length $L$ with $\chi^{(2)}$ nonlinearity, as illustrated in Fig.~\ref{fig1}(a).  The waveguide is pumped by a pair of counter-propagating fields at $2\omega_0$, exciting a pair of degenerate squeezing processes. Squeezed vacuum fields are generated and propagated in both directions, which amplify progressively along the waveguide ~\cite{PhysRevResearch.7.L012014}. The pump phases, $\overset{\rightarrow}{\theta}$ and $\overset{\leftarrow}{\theta}$, correspond to the left- and right- propagation directions, respectively. The waveguide modes are described by the field operator $A_k(z)$, which characterizes the amplitude and phase of the mode with wavevector $k$ at position $z$. The coupled-wave equations for the two propagating modes $k_1$ and $k_2$ in the waveguide are given by
	\begin{gather}
		\frac{dA_{k1}(z)}{dz}=-\frac{1}{2}\alpha _1A_{k1}(z)-\frac{i}{2}\mathcal{G} A_{k2}^{\ast}(z)e^{-i \Delta k  z},
		\notag \\
		\frac{dA_{k2}^{\ast}(z)}{dz}=-\frac{1}{2}\alpha _2A_{k2}^{\ast}(z)+\frac{i}{2}\mathcal{G} A_{k1}(z)e^{i \Delta k z},
	\end{gather}
	where $\alpha$ represents the loss in the system, and $\mathcal{G}$ is the parametric gain coefficient in both directions. We define the phase detuning as $\Delta k=k_1+k_2-2k_0$. In the absence of loss ($\alpha=0$), we solve the coupled-wave equations with the quantized electromagnetic field operator $ \overset{\rightarrow}{A}_k(z,t)=\overset{\rightarrow}{A}_k(z)e^{-i(\omega_kt-kz)}+\mathrm{c.c}$, and obtain:
	\begin{widetext}
		\begin{eqnarray}
			\overset{\rightarrow}{A}_{k1}\left( z \right) e^{i \Delta kz/2 }=\overset{\rightarrow}{A}_{k1}\left( 0 \right) \left[ \cosh \left( bz \right) -\frac{i\Delta k}{2b}\sinh \left( bz \right) \right]  -\frac{i\mathcal{G}}{2b}e^{i\vec{\theta}}\;\overset{\rightarrow}{A}\mkern-2mu^{\ast}_{k2}(0) \sinh \left( bz \right) ,
			\notag \\
			\overset{\rightarrow}{A}_{k2}\left( z \right) e^{i \Delta kz/2 }=\overset{\rightarrow}{A}_{k2}\!\left( 0 \right) \left[ \cosh \left( bz \right) +\frac{i\Delta k}{2b}\sinh \left( bz \right) \right]   -\frac{i\mathcal{G}}{2b}e^{i\vec{\theta}}\;\overset{\rightarrow}{A}\mkern-2mu^{\ast}_{k1}(0) \sinh \left( bz \right) ,
		\end{eqnarray}
	\end{widetext}
	where $b=\frac{1}{2}\sqrt{\mathcal{G} ^2-\left( \Delta k \right) ^2}$. The startpoint of the right (left) traveling-wave parametric mode is denoted as $\overset{\rightarrow}{A_k}(0)$ ($\overset{\leftarrow}{A_k}(L)$), and is assumed to be in its vacuum states. These modes satisfy the following commutation relation
	\begin{eqnarray}
		&[ \overset{\rightarrow}{A_k}\left( 0 \right) ,\overset{\rightarrow}{A}\mkern-2mu^{\dagger}_{k^\prime}(0) ] =\delta _{kk^\prime}, \quad  \langle \overset{\rightarrow}{A_k}\left( 0 \right) \overset{\rightarrow}{A}\mkern-2mu^{\dagger}_{k}(0) \rangle =1, \notag \\
		&[ \overset{\leftarrow}{A_k}\left( L \right) ,\overset{\leftarrow}{A}\mkern-2mu^{\dagger}_{k^\prime}(L) ]=\delta _{kk^\prime}, \quad \langle \overset{\leftarrow}{A_k}\left( L \right) \overset{\leftarrow}{A}\mkern-2mu^{\dagger}_{k}(L)\rangle  =1.\notag
	\end{eqnarray}
	By imposing the phase-matching condition $\Delta k=0$, we derive the field operators as
	\begin{eqnarray}
		\!\!\overset{\rightarrow}{A}_k \!\left( z \right)\!=\!\overset{\rightarrow}{A}_k \!\left( 0 \right)\! \cosh \!\left( \! 
		\frac{\mathcal{G}}{2}d^R \! \right)\!\!-\! 
		ie^{i\overset{\rightarrow}{\theta}}
		\overset{\rightarrow}{A}\mkern-2mu^{\ast}_{2k_0-k}(0) \!\sinh\! \left( \! 
		\frac{\mathcal{G}}{2}d^R \! \right)\!,	
		\notag\\
		\!\!\overset{\leftarrow}{A}_k \!\left( z \right)\! =\! 
		\overset{\leftarrow}{A}_k \!\left(L \right)\! \cosh \! \left( \! 
		\frac{\mathcal{G}}{2} d^L \! \right)\!\!-\! 
		ie^{i\overset{\leftarrow}{\theta}}\overset{\leftarrow}{A}\mkern-2mu^{\ast}_{-2k_0-k}\!(L)
		\!\sinh \! \left(\! \frac{\mathcal{G}}{2} d^L \! \right)\!\!,
		\label{field_ak}
	\end{eqnarray}
	with $d^R=z$ ($d^L=L-z$).
	
	Assuming that the coupling strength at position $z$ is $g(z)$, the interaction Hamiltonians for the right- and left-propagating directions are written as
	\begin{gather}
		H^s_{R}=g\left( z \right) \int_0^{2k_0}{dk\overset{\rightarrow}{A}_k\left( z \right) e^{ikz}e^{i\delta _kt}\sigma _+}+\mathrm{H}.\mathrm{c}.,
		\notag \\
		H^s_{L}=g\left( z \right) \int_{-2k_0}^0{dk\overset{\leftarrow}{A}_k\left( z \right) e^{ikz}e^{i\delta _kt}\sigma _+}+\mathrm{H}.\mathrm{c}.,
		\label{H_left and right}
	\end{gather}
	where $\delta_k=\omega_k-\omega_0$ denotes the detuning, and $\sigma_+= |e\rangle \langle g|$ is the atomic transition operator. By substituting Eq.~(\ref{field_ak}) into (\ref{H_left and right}), the Hamiltonian is rewritten as
	\begin{gather}
		H_R^s=g\left( z_0 \right) \int_0^{\infty}dk{S_R^{\dagger}\left( z_0 \right) e^{ikz_0}e^{i\delta _kt}\overset{\rightarrow}{A}_k\left( 0 \right)}  +\mathrm{H}.\mathrm{c}., \notag \\
		H_L^s=g\left( z_0 \right)\int_{-\infty}^0{dkS_L^{\dagger}(z_0) e^{ikz_0}e^{i\delta _kt}\overset{\leftarrow}{A}_k\left( L \right) +\mathrm{H}.\mathrm{c}.}
	\end{gather}
	
	We then extend this formalism to the case of $N$ giant atoms, labeled by $n=1,2, \cdots, N $. Each atom $n$ is nonlocally coupled to the parametric waveguide at $M$ equidistant points, denoted $z_{n i}$ with $i=1,2, \cdots, M $. The corresponding Hamiltonian takes the form
	\begin{gather}
		H=\frac{1}{2}\sum_{n}{\Delta_n \sigma_z^n}+H_{\mathrm{int}}, \\
		H_{\mathrm{int}}=\sum_{n,i}{g\left( z_{n i} \right) \int_0^{\infty}dk{S_R^{\dagger}\left( z_{n i} \right) e^{ikz_{n i}}e^{i\delta _kt}\overset{\rightarrow}{A}_k\left( 0 \right)}} \notag \\
		+\!\sum_{n,i}{g\left( z_{n i} \right)\int_{-\infty}^0\!{dkS_L^{\dagger}(z_{n i}) e^{ikz_{n i}}e^{i\delta _kt}\overset{\leftarrow}{A}_k\left( L \right)}\! +\!\mathrm{H}.\mathrm{c}.},
	\end{gather}
	where $\Delta_n=\omega_n-\omega_0$. The quantum-jump operators $S_{R(L)}\left( z_{n i} \right)$ describe the coupling of atomic transitions to right- (left-) propagating waveguide modes, i.e.,
	\begin{eqnarray}
		\! S_R\!\left( z_{n i} \right)\!=\!\cosh\! \left(\! \frac{\mathcal{G} 
		}{2}d^R_{n i}\! \right)\! \sigma _-^n \!\!-\! 
		ie^{i\overset{\rightarrow}{\theta}} \! \!
		\sinh\! \left(\! \frac{\mathcal{G} }{2}d^R_{n i} \!\right)\! 
		\!e^{2ik_0z_{n i}}\sigma _+^n,
		\label{S_right}
		\\
		\! S_L\!\left( z_{n i} \right)\!=\!\cosh \!\left(\! \frac{\mathcal{G} 
		}{2}d^L_{n i}\! \right)\! \sigma _-^n \!\!-\! 
		ie^{i\overset{\leftarrow}{\theta}}\!\!
		\sinh \!\left(\! \frac{\mathcal{G} }{2}d^L_{n i}\! \right)\! 
		\!e^{\!-2ik_0z_{n i}}\sigma _+^n.
		\label{S_left}
	\end{eqnarray}
	where the parametric gain $\mathcal{G}$ depends on the second-order nonlinear susceptibility $\chi^{(2)}$ and the power of the pump field, i.e., $ \mathcal{G} \propto \chi ^{\left( 2 \right)}A_{2k_0}\left( 0 \right) $. 
	\subsection{Waveguide-mediated interactions between giant atoms}
	\begin{figure*}
		\centering
		\includegraphics[width=0.9\linewidth]{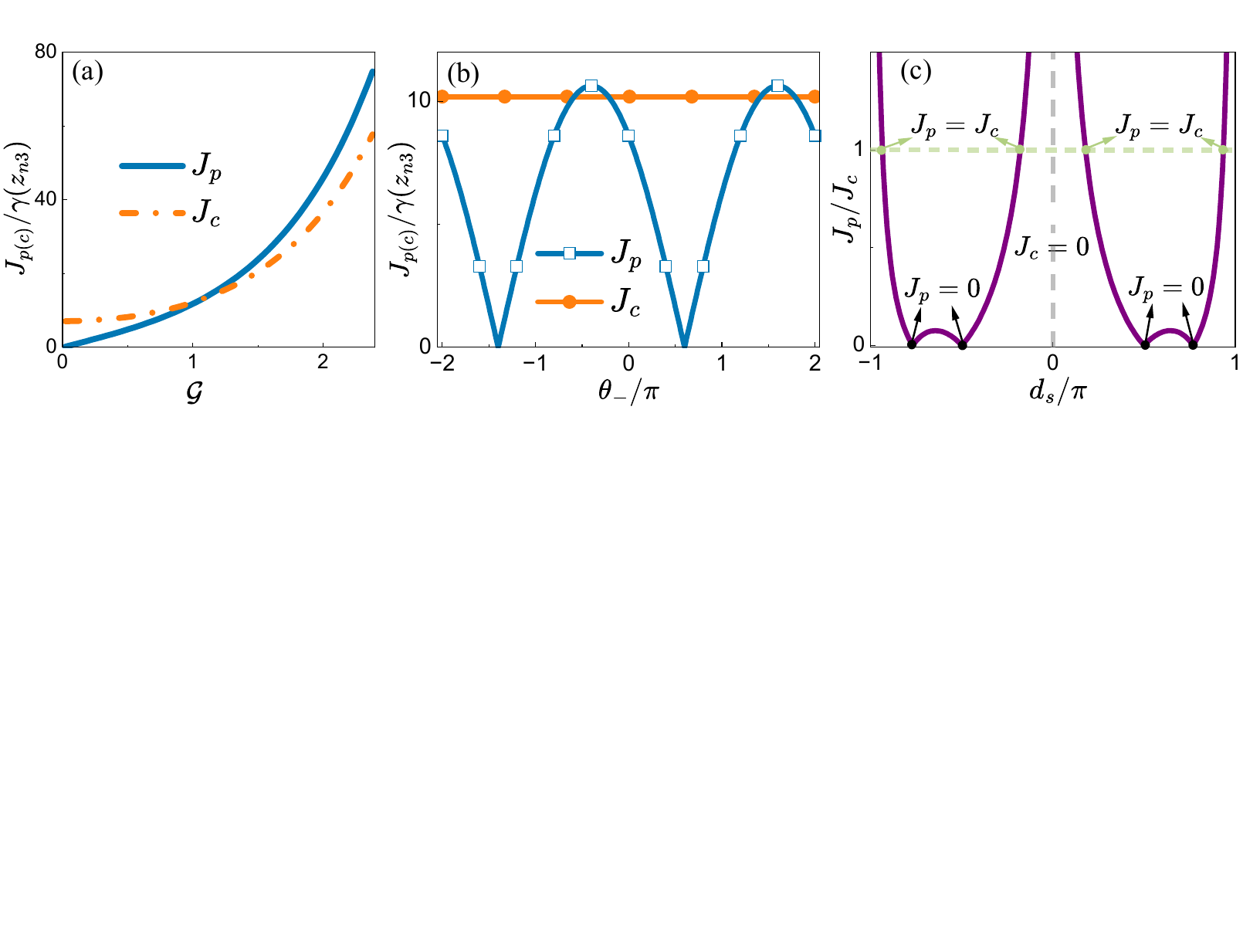}
		\caption{Dependence of pairing and exchange interaction strengths on (a) parametric gain $\mathcal{G}$, with fixed phase difference $\theta_- = 0$ and atomic spacing $d_s = 0.2\pi$; and (b) phase difference $\theta_-$, with fixed parametric gain $\mathcal{G} = 0.8$ and atomic spacing $d_s = 0.2\pi$. (c) The ratio $J_p/J_c$ as a function of atomic spacing $d_s$, with parametric gain $\mathcal{G} = 0.8$ and phase difference $\theta_- = 0$. The black dots indicate $J_p=0$. The vertical dashed line marks $J_c=0$, and the horizontal dashed lines denote $J_p=J_c$.}
		\label{fig2}
	\end{figure*}
	
	 To identify conditions under which the system is immune to decoherence arising from the parametric amplification process, we employ the SLH formalism to derive the effective quantum jump operators and determine the parameter regimes which enable decoherence-free interaction. For multi-atom platforms, we compute the SLH triplet $G=(\mathcal{S},\mathcal{L},H)$ by cascading the individual atom-waveguide interactions~\cite{PhysRevLett.120.140404,Combes04052017,Gough_2009}. In the following, we focus on the configuration with two giant atoms spaced by $d_s$,
	and $\varphi_1=k_0(z_{b_i}-z_{a_i})$ and $\varphi_2=k_0(z_{a_{i+1}}-z_{b_i})$ denote the phase differences between adjacent coupling points, with $\varphi_1+\varphi_2=\pi$ [see Fig.~\ref{fig1}(b)]. The total triplet, combining both propagation directions, is given by
	\begin{eqnarray}
		G_{tot}\!=\! G_R\boxplus G_L \!=\!\left( \left[ \begin{matrix}
			\mathcal{S}_{\mathbf{R}}&		0\\
			0&		\mathcal{S}_{\mathbf{L}}\\
		\end{matrix} \right] ,\left[ \begin{array}{c}
			\mathcal{L} _{\mathbf{R}}\\
			\mathcal{L} _{\mathbf{L}}\\
		\end{array} \right] ,H_{R}\!+\! H_{L} \right) .
	\end{eqnarray}
	The jump operators $\mathcal{L}_{R(L)}$, which account for phase contributions from all coupling points, are written as (see Appendix \ref{appendix_a})
	\begin{gather}
		\mathcal{L} _{R\left( L \right)}=\sum_{n ,i}{\sqrt{\frac{\gamma _{n ,i}}{2}}e^{i\phi _{n i}^{R(L)}}S_{R\left( L \right)}(z_{n i})},
		\label{quantum jump}
	\end{gather}
	where $\gamma _{n,i}$ represents the relaxation rates of atom $n$ coupling to point $i$, given by $\gamma _{n,i}=\sqrt{2\pi g^2\left( z_{n i} \right)/v_n} $, and $v_n$ denotes the group velocity of the waveguide (we set $v_n=1$). Here, $\phi_{n i}^{R(L)}$ represents the phase difference between the $i$-th coupling points of atom $n$ and the reference point. For the right-propagating modes the phase reference is taken at the first coupling point $z_{a1}$, while for the left-propagating modes it is referenced to $z_{bM}$, as depicted in Fig.~\ref{fig1}(b). Note that for small atoms ($M=1$), the decoherence caused by quantum jump $\mathcal{L}_{R(L)}$ will be significantly enhanced by the squeezing process.
	
	To eliminate the noise introduced by the parametric squeezing noise in the waveguide, we engineer the setup to achieve vanishing jump operators $\mathcal{L}_R=\mathcal{L}_L=0$ through destructive interference of dissipation channels. For the case where the giant atoms coupled to the waveguide at two coupling points ($M=2$), the solution requires
	\begin{gather}
		\frac{g\left( z_{n1} \right)}{g\left( z_{n2} \right)}\!=\!\frac{\cosh^2 \left( \frac{\mathcal{G} }{2}z_{n2} \right)}{\cosh^2 \left( \frac{\mathcal{G} }{2} z_{n1}\right)},\quad \frac{g\left( z_{n1} \right)}{g\left( z_{n2} \right)}\!=\!\frac{\sinh^2 \left( \frac{\mathcal{G} }{2}z_{n2} \right)}{\sinh^2 \left( \frac{\mathcal{G} }{2} z_{n1}\right)}.
	\end{gather}
	This demonstrates that under real-valued parameters, the system cannot simultaneously satisfy the given coupling strength ratio condition when $M=2$ ($z_{n 1} \ne z_{n 2}$). Therefore, each giant atom must interact with the waveguide through at least three coupling points ($M\geqslant 3$) to satisfy the totally destructive interference. We take three coupling points, and set $g\left( z_{n3} \right)$ as the reference coupling strength. The vanishing jump operator constraint dictates that the coupling strengths must satisfy the following relations
	\begin{gather}
		\frac{g\left( z_{n 1} \right)}{g(z_{n 3})} =1, \;\;\;\;\;
		\frac{g\left( z_{n 2} \right)}{g(z_{n 3})} =4\cosh^2 \left[ \frac{\mathcal{G}}{2}\left( z_{n 2}-z_{n 1} \right) \right] .
	\end{gather}
	Intriguingly, there exists enhanced coherent exchange and pairing interactions between giant atoms, even when the aforementioned decoherence process is cancelled by the interference. To simplify the form of the Hamiltonian, we first introduce the phase difference $\theta_{\pm}=\overset{\rightarrow}{\theta}\pm\overset{\leftarrow}{\theta}$ and apply the gauge phase transformation $ \sigma _{+}^{a,b}=e^{-i \theta _+/2 }\sigma _{+}^{a,b}$. Under this transformation, the coherent interaction strength and Hamiltonian are respectively derived as
	\begin{widetext}
		\begin{eqnarray}
			H &=&\frac{\Delta_a}{2}\sigma _{z}^{a}+\frac{\Delta_b}{2}\sigma _{z}^{b} 
			+J_p\sigma _{+}^{a}\sigma _{+}^{b}+J_c\sigma _{+}^{a}\sigma _{-}^{b}+\mathrm{H}.\mathrm{c}.,\\
			J_c &=&\sum_{i,j}{\sqrt{\frac{\pi g(z_{ai})g(z_{bj})}{2}}\sin |z_{bj}- z_{ai}| \cosh \left[ \frac{\mathcal{G}}{2}\left( z_{bj}-z_{ai} \right) \right]},\\
			J_p &=&\sum_{i,j}{\sqrt{\frac{\pi g(z_{ai})g(z_{bj})}{2}} \mathrm{sgn}\left( z_{bj}- z_{ai} \right) }  
			\cos \left[ \frac{\theta _-}{2}+ z_{ai}+ z_{bj} \right]\sinh \left[ \frac{\mathcal{G}}{2}\left( z_{bj}- z_{ai} \right) \right],
			\label{J_p}	
		\end{eqnarray}
	\end{widetext}
	where $J_p$ and $J_c$ quantify parametric pairing and coherent exchange interactions, respectively. The pairing interaction strength $J_p$ term is mediated by the squeezed vacuum fluctuations in the $\chi^{(2)}$ waveguide, enabling a parametric process through the simultaneous excitation (de-excitation) of two atoms. In contrast, $J_c$ describes the coherent transfer of excitation between atoms through exchanging photons via the waveguide. This dual structure reflects the interplay between squeezing-induced nonlinearity and conventional dipole-dipole coupling.
	
	As shown in Fig.~\ref{fig2}(a), the decoherence-free interaction strengths are significantly enhanced by the squeezing process when increasing $\mathcal{G}$. Moreover, our proposal offers two independent parameters, the phase difference $\theta_-$ between the counter-propagating pump fields, and the distance $d_s$ between two giant atoms [see Fig.~\ref{fig2}(b,c)]. This dual-control capability allows for dynamic modulation of the relative strength $J_p/J_c$ over a wide range, enabling the realization of distinct quantum regimes. Such flexibility is particularly valuable for quantum simulation, where different interaction types are required to emulate various many-body Hamiltonian.
	\begin{figure}
		\centering
		\includegraphics[width=0.85\linewidth]{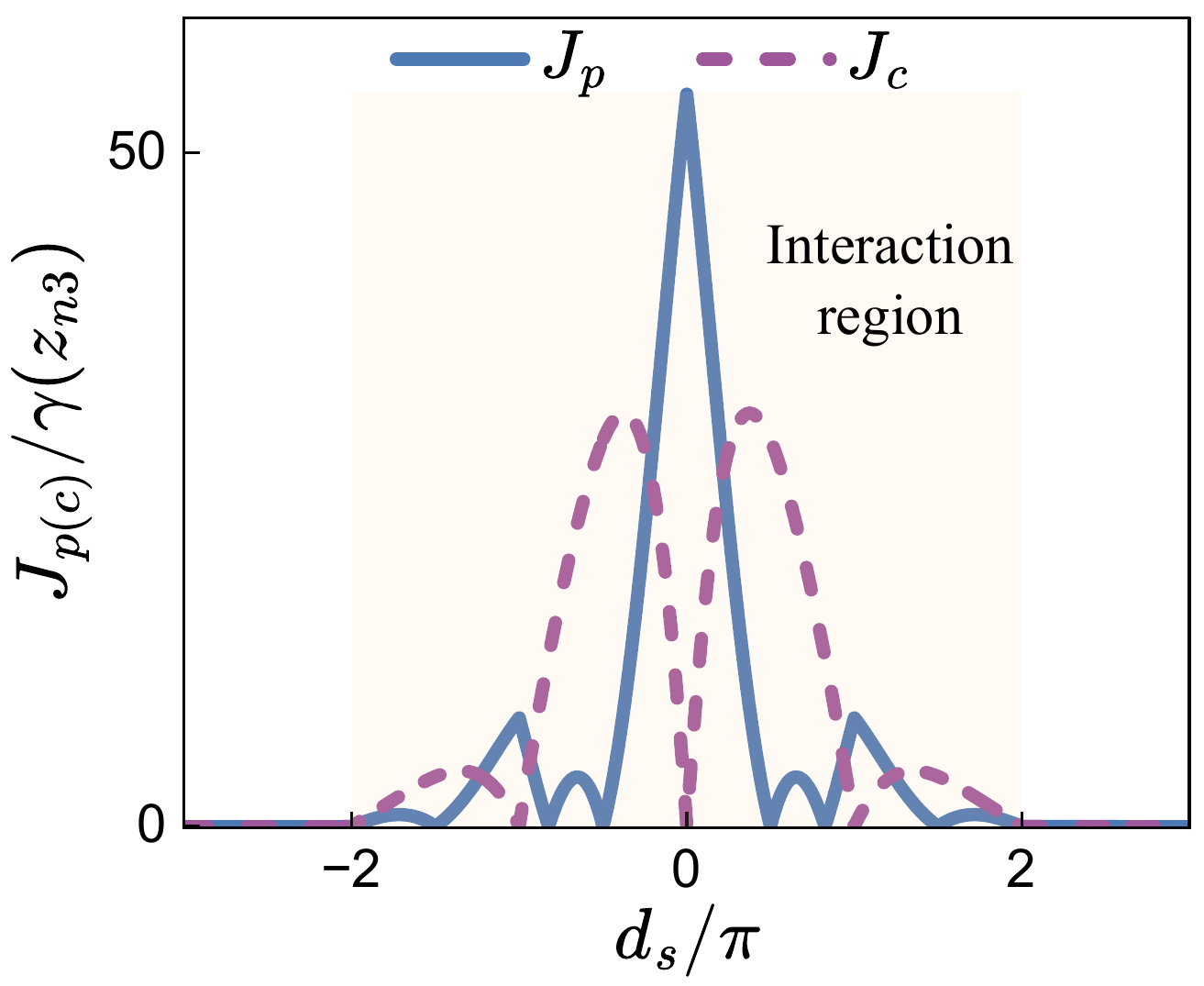}
		\caption{Pairing and exchange interaction strength versus the distance $d_s$, with $\mathcal{G}=1.2$ and $\theta_-=0.$}
		\label{fig3}
	\end{figure}
	
	 In Fig.~\ref{fig3}, the overlapping coupling regions adopt a braided configuration analogous to that in Ref.~\cite{PhysRevLett.120.140404}, which ensures that coupling occurs only between atoms separated by a distance less than $2\pi$ (i.e., nearest neighbor), while interactions between  next-nearest-neighbor atoms vanish due to the absence of overlapping coupling regions, thereby enabling giant atoms to exhibit strictly nearest-neighbor couplings~\cite{PhysRevApplied.10.054062,PhysRevApplied.14.064016}. This feature is critical to simulate various condensed matter models, as many quantum simulation platforms inherently suffer from unavoidable next-nearest-neighbor couplings, the elimination of which typically poses significant challenges~\cite{PhysRevApplied.10.054062,PhysRevApplied.14.064016}. Remarkably, our proposal intrinsically overcomes this limitation through the tunable spatial structure.
	\begin{figure}
		\centering
		\includegraphics[width=1\linewidth]{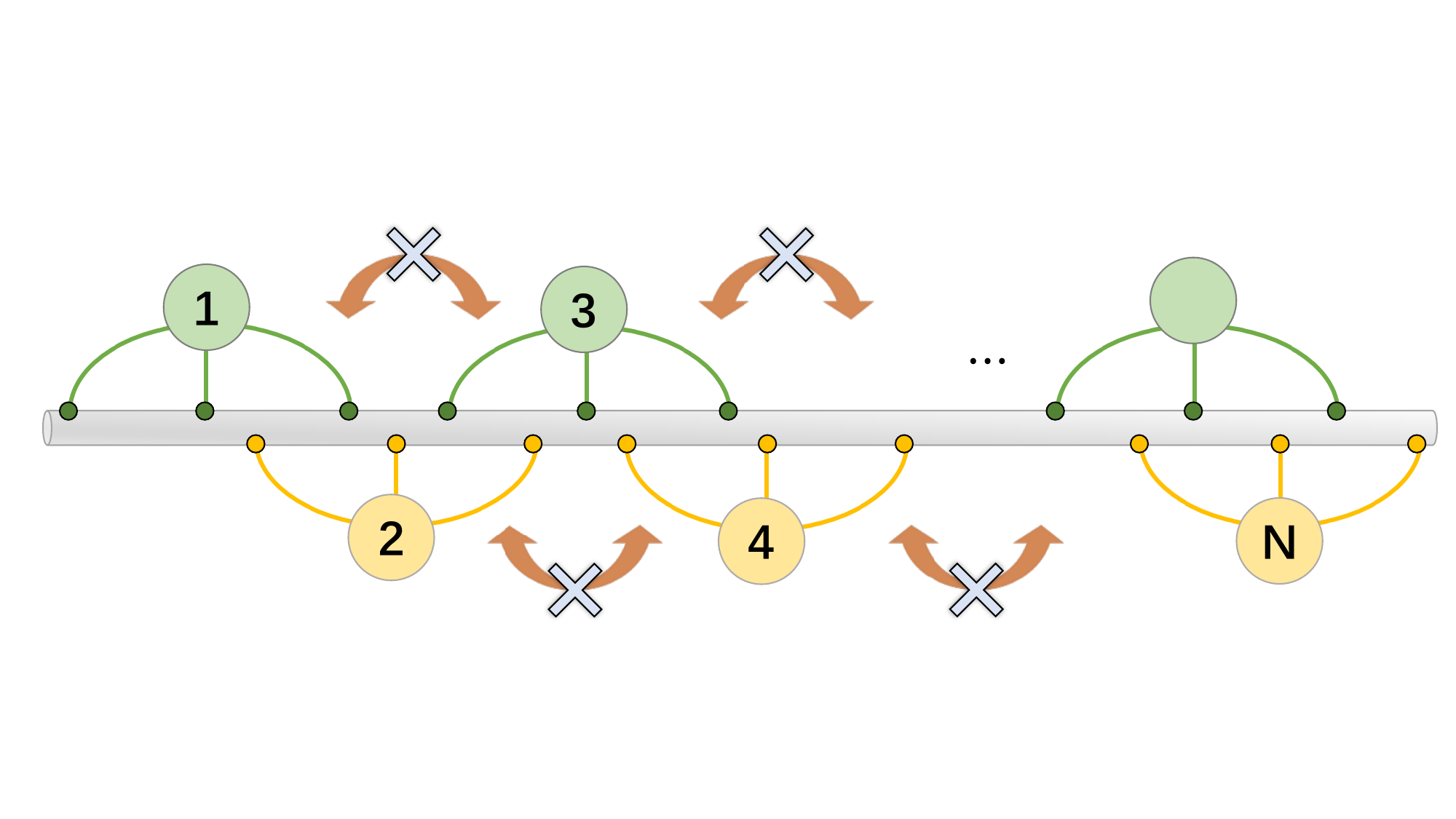}
		\caption{A chain of $N$ equally spaced giant atoms coupled to a parametric waveguide, with no next-nearest-neighbor interactions in the chain.}
		\label{fig4}
	\end{figure}
	
	\section{Many-body quantum simulation with giant atoms}
	\begin{figure*}
		\centering
		\includegraphics[width=1\linewidth]{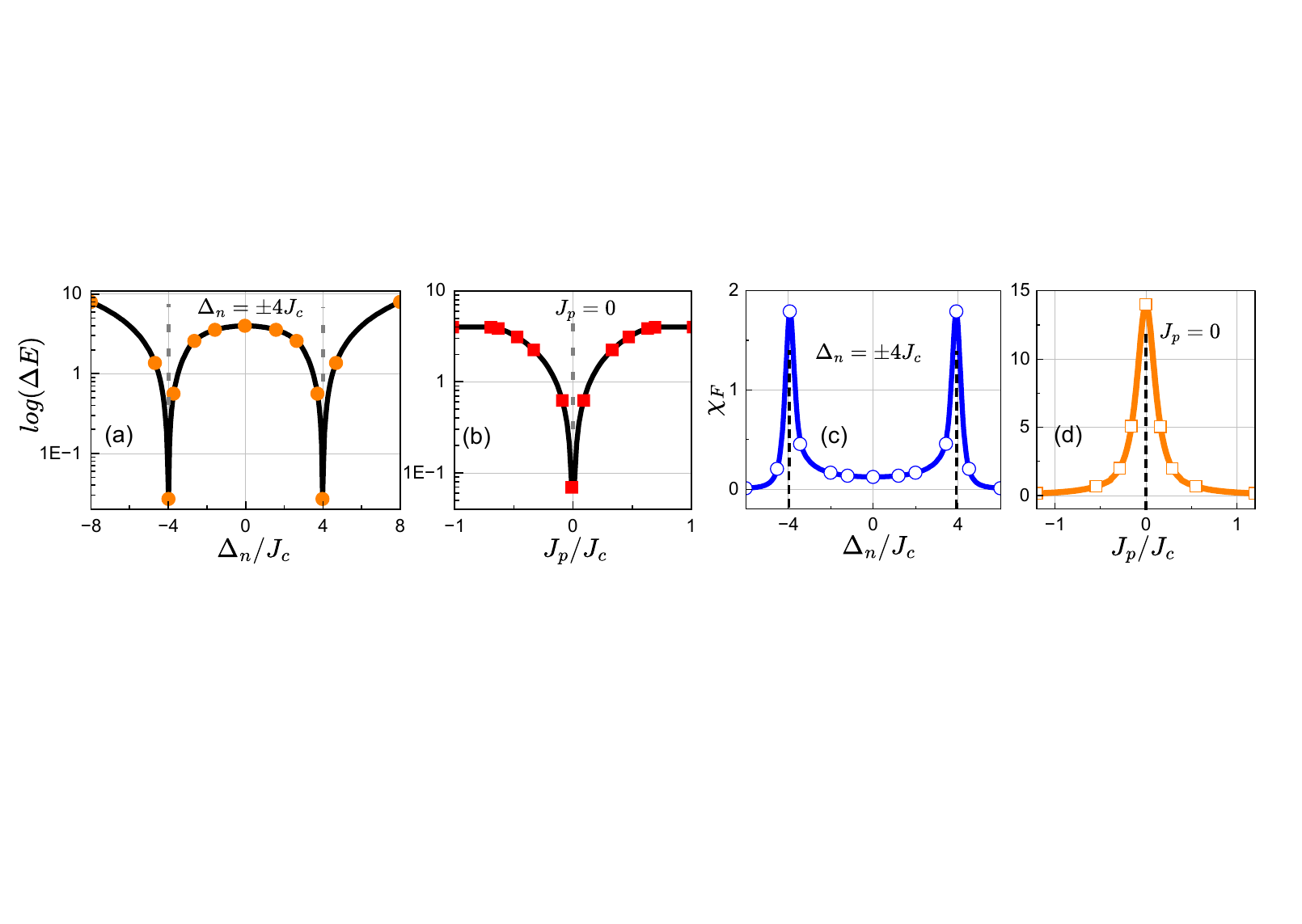}
		\caption{
			(a) Energy gap as a function of the giant atom frequency $\Delta_n$, with $J_p = 0.5$.  
			(b) Energy gap as a function of the pairing interaction strength $J_p$, with $\Delta_n = 2.0$.  
			(c) Fidelity susceptibility of the XY model versus atomic detuning $\Delta_n$, exhibiting pronounced peaks at the critical points $\Delta_n = \pm 4J_c$, where $J_p = 0.5$ and $\delta = 0.01$.  
			(d) Fidelity susceptibility as a function of the pairing interaction strength $J_p$, for $\Delta_n = 2.0$ and $\delta = 0.01$.  
			All results are obtained for a finite system of size $N = 16$, with the exchange interaction strength set to $J_c = 1$.
		}
		\label{fig5}
	\end{figure*}
	The unique design of the braided giant atoms, where each atom couples to a waveguide at multiple and non-adjacent points, not only eliminates unwanted next-nearest-neighbor interactions but also enables precise control over tunable pairing and exchange couplings. This combination of features makes our platform particularly well-suited for simulating paradigmatic many-body quantum models, such as the Kitaev chain and anisotropic XY model, which exhibit rich phenomena including topological phases and quantum phase transitions.
	
	To explore the many-body physics, we begin by mapping our system of $N$ giant atoms coupled to a parametric waveguide (see Fig.~\ref{fig4}) onto an effective spin model. By expressing the atomic transitions in terms of Pauli operators and applying a spin rotation $ \sigma _{\pm}=\sigma _x\pm i\sigma _y$, we find that the engineered couplings yield purely real exchange ($J_c$) and pairing ($J_p$) interaction strengths, i.e., $J_p=J_{p}^{\ast} $ and $J_c=J_{c}^{\ast}$. Therefore, this condition ensures that the Hamiltonian can be written in the standard form of the anisotropic XY model as
	\begin{align}
		H =&\; 2\sum_{n=1}^{N-1} \Big[ 
		\left( J_p + J_c \right) \sigma_{x}^{n}\sigma_{x}^{n+1} \notag \\
		& + \left( J_c - J_p \right) \sigma_{y}^{n}\sigma_{y}^{n+1}
		\Big] 
		+ \sum_{n=1}^N \Delta_n \sigma_{z}^{n}.
	\end{align}
	
	To analyze topological or critical behavior, we apply the Jordan-Wigner transformation, which maps spin operators to fermionic operators
	\begin{gather}
		\sigma _{j}^{z}=1-2c_{j}^{\dagger}c_j, \notag\\
		\sigma _{j}^{x}=-\prod_{i=1}^{j-1}{\left( 1-2c_{i}^{\dagger}c_i \right) \left( c_{j}^{\dagger}+c_j \right)},
		\notag\\
		\sigma _{j}^{y}=-i\prod_{i=1}^{j-1}{\left( 1-2c_{i}^{\dagger}c_i \right) \left( c_{j}^{\dagger}-c_j \right)}.
	\end{gather}
	The corresponding many-body Hamiltonian is given by
	\begin{eqnarray}
		&&H=H_{\Delta}+H_I+H_{B},\notag \\
		&&H_{\Delta}=-\sum_{n=1}^N{\Delta_n \left( 2c_{n}^{\dagger}c_n - 1 \right)}, \notag \\
		&&H_I=\sum_{n=1}^{N-1}{J_cc_{n+1}^{\dagger}c_n - J_pc_{n+1}^{\dagger}c_{n}^{\dagger}+\mathrm{H}.\mathrm{c}.},\notag \\
		&&H_B=-\left( -1 \right) ^{N_f}\left( J_c c_{N}^{\dagger}c_1-J_p c_{1}^{\dagger}c_{N}^{\dagger}+\mathrm{H}.\mathrm{c}. \right),
	\end{eqnarray}
	where the last term enforces periodic boundary conditions. Here,  $N_f=\sum\nolimits_{n=1}^N{c_{n}^{\dagger}c_n}$ denotes the total fermion number. 
	
	Then, we move to momentum space via the Fourier transform $ c_n=\sum_k{e^{ikn}c_k}/\sqrt{N}$. For even system size and appropriate fermion parity, the allowed momenta are quantized as $k=\left( 2m+1 \right) \pi /N $ for even $N_f$ and $k=2m\pi/N$ for odd $N_f$~\cite{PhysRevB.98.184415,Yang_2019,Dutta_Aeppli_Chakrabarti_Divakaran_Rosenbaum_Sen_2015}. Therefore, the Hamiltonian in momentum space is written as~\cite{Jafari_2019,Deng_2008}
	\begin{align}
		H_k \!=\! \left(16J_c\cos k \! - \! 2\Delta_n\right) c_{k}^{\dagger}c_k  
		\!-\! 8iJ_p\sin k \left(c_{k}^{\dagger}c_{-k}^{\dagger}\! +\! c_k c_{-k}\right).
	\end{align}
	Through a Bogoliubov transformation with optimally selected $\theta_k$, the off-diagonal terms in the Hamiltonian can be effectively suppressed~\cite{Dutta_Aeppli_Chakrabarti_Divakaran_Rosenbaum_Sen_2015,PhysRevA.94.022112}. The transformation is defined as
	\begin{gather}
		d_{k}^{\dagger}=\sin \theta _kc_k+i\cos \theta _kc_{-k}^{\dagger},
		\notag	\\
		d_{-k}^{\dagger}=\sin \theta _kc_{-k}-i\cos \theta _kc_{k}^{\dagger}.
	\end{gather}
	where the angle $\theta_k$ satisfies the relation $$\tan \left( 2\theta _k \right) =\frac{-J_p\sin \left( k \right)}{J_c\cos \left( k \right) +\Delta_n}. $$ Then, the Hamiltonian is diagonalized as
	\begin{gather}
		H=\sum_{k>0}{\epsilon _k\left( d_{k}^{\dagger}d_k+d_{-k}^{\dagger}d_{-k}-1 \right)}.
	\end{gather}
	
	By solving the Hamiltonian in momentum space, the excitation spectrum of the system is derived as~\cite{Saghafi_2019}
	\begin{eqnarray}
		\frac{\epsilon_k^2}{4}  &=& \Delta_n^2 + 8(J_p^2 + J_c^2) 
		- 8\Delta_n J_c \cos k \notag \\
		&-& 8(J_p^2 - J_c^2) \cos 2k.
		\label{dispersion_relation}
	\end{eqnarray}
	\begin{figure}
		\centering
		\includegraphics[width=0.6\linewidth]{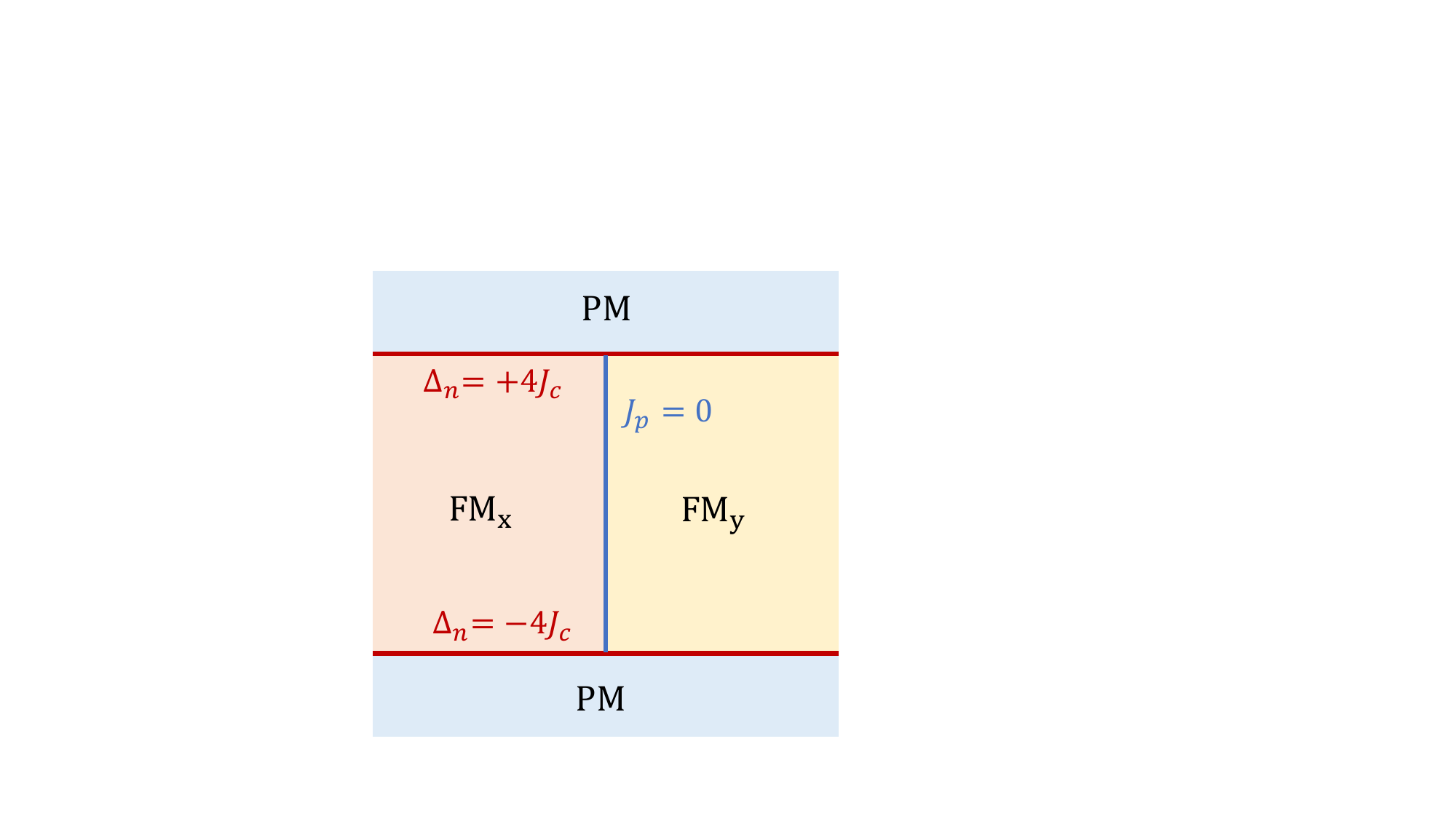}
		\caption{ The phase diagram in the plane of the pairing interaction strength $J_p$ and atomic frequency $\Delta_n$.}
		\label{fig6}
	\end{figure}
	 In gapped quantum phases, the energy gap $\Delta E =\mathrm{min}_k \epsilon_k$ protects the ground state from local perturbations. Crucially, a vanishing gap ($\Delta E \to 0$) signals a quantum phase transition. Thus, tracking the gap as a function of $\Delta_n$ and $J_p$ provides a direct probe of criticality. Fig.~\ref{fig5}(a,b) shows how the energy gap evolves with atomic detuning $\Delta_n$ and pairing interaction strength $J_p$. Notably, the gap closes at $J_p=0$ and $\Delta_n=\pm 4 J_c$, consistent with known critical points of the transverse-field XY model~\cite{Dutta_Aeppli_Chakrabarti_Divakaran_Rosenbaum_Sen_2015, Suzuki_2013}, confirming that our engineered giant atom–waveguide system realizes the universal physics of this paradigmatic quantum critical model. A comprehensive phase diagram in the parameter space is shown in Fig.~\ref{fig6}, delineating the distinct quantum phases and their boundaries.
	
	However, in finite-size systems the excitation gap never strictly reaches zero due to finite-size effects, which can obscure the precise location of the critical point. To overcome this limitation, we employ quantum fidelity, which is a measure of the overlap between ground states at nearby parameter values. Specifically, we compute the fidelity susceptibility $\chi_F$ defined as
	\begin{eqnarray}
		F(h,h+\delta h)&=&|\langle\psi_0(h)|\psi_0(h+\delta h)\rangle|,\\
		\chi _F\left( h \right)& =&\underset{\delta \rightarrow 0}{\lim}\frac{-2\ln F}{\delta h^2}.
	\end{eqnarray}
	where $h$ denotes a control parameter, i.e., $\Delta_n$ and $J_p$. The fidelity susceptibility exhibits a pronounced peak at the quantum critical point, as the ground state changes most rapidly in its vicinity. As shown in Fig.~\ref{fig5}(c,d), the fidelity susceptibility shows sharp peaks at $\Delta_n=\pm 4 J_c$ and $J_p=0$, in excellent agreement with the gap-closing condition. This confirmation establishes the presence of quantum phase transitions in our platform via both spectral and ground-state properties. Moreover, the system can be continuously driven across distinct quantum phases by tuning the pump phase difference $\theta_-$, which controls the pairing amplitude $J_p$ [cf. Eq.~(\ref{J_p})]. This tunability enables precise simulation of many-body quantum phase transitions, offering a versatile platform for studying quantum critical phenomena. Importantly, while our model is implemented with qubits, it can be extended to higher-dimensional paradigms such as qutrits, enabling the emulation of spin-1 chains and other exotic many-body phases \cite{PhysRevA.106.033705}. This scalability demonstrates the versatility of our proposed platform in exploring correlated quantum dynamics and high-dimensional entanglement in engineered atomic platforms.
	
	\section{Experimental Realization}
	\begin{figure*}
		\centering
		\includegraphics[width=0.98\linewidth]{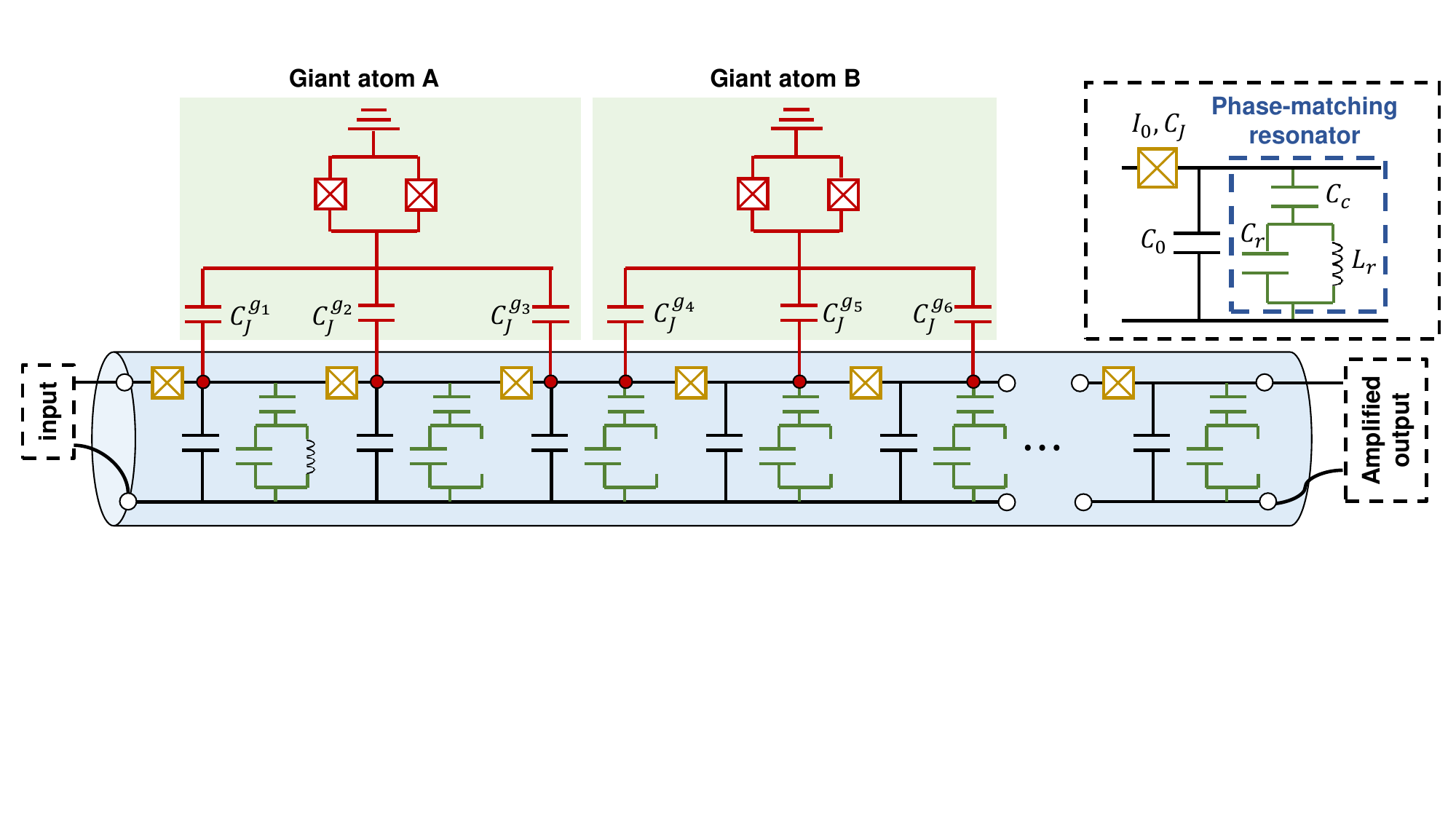}
		\caption{Circuit diagram of giant atoms with three coupling points interacting with a Josephson traveling-wave parametric amplifier. The JTWPA is designed as a lumped-element transmission line, consisting of unit cells with Josephson junctions (orange), shunt capacitors, and phase-matching resonators (green).  }
		\label{fig7}
	\end{figure*}
	As illustrated in Fig.~\ref{fig7}, we present the experimental circuit 
	diagram of the interaction using superconducting circuits. In our model, 
	the waveguide is implemented using a Josephson traveling-wave parametric 
	amplifier (JTWPA).  The giant atoms are implemented using transmon qubits 
	which couple to the waveguide via a capacitor 
	\cite{PhysRevLett.128.223602,PhysRevLett.126.043602}. The coupling 
	strength $g_k$ takes the 
	form~\cite{PhysRevLett.126.043602,PhysRevA.109.013716}
	\begin{gather}
		g_k=\frac{e}{\hbar}\frac{C_{J}^{g}}{C_{\Sigma}}\sqrt{\frac{\hbar \omega _k}{C_W}},
	\end{gather}
	where $ C_{\Sigma} $ denotes the total capacitance of the atom, with $ C_W $ denoting the total capacitance of the waveguide. Therefore, in practice, the coupling strength can be tuned by engineering the capacitance or by adjusting the waveguide properties.
	
	The JTWPA is schematically depicted as a lumped-element transmission line comprising unit cells with Josephson junctions, shunt capacitors, and phase-matching resonators (in the blue rectangle) \cite{Macklin_2015,PhysRevApplied.12.044051,Rizvanov_2024,PhysRevLett.113.157001,White_2015}. In experimental setups, the parametric gain $\mathcal{G}$ characterizes the strength of the squeezing generated by a JTWPA, and is quantified by the microwave power gain, i.e., the ratio of output to input signal power under strong pump drive \cite{Macklin_2015, PhysRevApplied.12.044051}.
	
	Based on the circuit diagram in Fig.~\ref{fig7}, we derive the coupled wave equations for the JTWPA by assuming traveling wave solutions and neglecting dissipation. Under the undepleted pump approximation, these equations describe the energy exchange between the pump, signal, and idler, are expressed as ~\cite{PhysRevLett.113.157001}
	\begin{subequations}
		\begin{align}
			\frac{\partial a_s}{\partial x} - i \kappa_s a_i^* e^{i(\Delta k_L + 2\alpha_p - \alpha_s - \alpha_i)x} &= 0, \\
			\frac{\partial a_i}{\partial x} - i \kappa_i a_s^* e^{i(\Delta k_L + 2\alpha_p - \alpha_s - \alpha_i)x} &= 0.
		\end{align}
	\end{subequations}
	The terms $a_s$ and $a_i$ in the above equations represent the signal and idler amplitudes. The coupling factors $\alpha_p$, $\alpha_s$ and $\alpha_i$, which describe the wavevector shifts induced by pump self- and cross-phase modulation, are given by ~\cite{PhysRevLett.113.157001}
	\begin{align}
		&\alpha_{s(i)} = \frac{2\kappa k_{s(i)}^3 a^2 i Z_2(\omega_{s(i)})}{L \omega_{s(i)}}, \notag \\  &\kappa_{s(i)} = \frac{\kappa(2k_p - k_{i(s)}) k_s k_i i Z_2(\omega_{s(i)}) a^2}{L \omega_{s(i)}}, \notag\\
		&\alpha_p = \frac{\kappa k_p^3 a^2 i Z_2(\omega_p)}{L \omega_p}, \quad  \kappa = \frac{a^2 k_p^2 |Z_{\text{char}}|^2}{16 L^2 \omega_p^2} \left( \frac{I_p}{I_0} \right)^2.
	\end{align}
	The solutions are
	\begin{subequations}
		\begin{gather}
			a_s(x) = \Big\{ a_s(0) \Big[ \cosh(bx) - \frac{i\Delta k}{2b} \sinh(bx) \Big] \notag\\
			\qquad + \frac{i\kappa_s}{b} a_i^*(0) \sinh(bx) \Big\} e^{i\Delta k x / 2}, \\
			a_i(x) = \Big\{ a_i(0) \Big[ \cosh(bx) - \frac{i\Delta k}{2b} \sinh(bx) \Big] \notag\\
			\qquad + \frac{i\kappa_i}{b} a_s^*(0) \sinh(bx) \Big\} e^{i\Delta k x / 2}.
		\end{gather}
	\end{subequations}
	where $b=\sqrt{\kappa _s\kappa _{i}^{\ast}-\left( \Delta k/2 \right) ^2}$ represents the gain coefficient.
	
	Recent experiments with giant atoms have indicated that both equal and engineered unequal coupling strengths can be realized with high precision using current circuit-QED fabrication techniques~\cite{Nature_superconducting, kannan2023ondemand}. These studies further indicate that small deviations from ideal coupling conditions induce only negligible decoherence and residual interactions, well below the scale of coherent dynamics. This confirms that the decoherence-free regime is robust against typical experimental imperfections, supporting the practical realizability of our proposal.
	\section{Conclusions} 
	In this work, we propose a novel approach to realizing amplified decoherence-free many-body interactions with giant atoms coupled to a parametric waveguide. By exploiting destructive interference between different coupling points, the system effectively eliminates strong decoherence caused by squeezing noise without any additional overheads. Unlike cavity-based paradigms, our implementation employs a scalable traveling-wave parametric waveguide, making it suitable for simulating many-body physics. The proposed setup supports both pairing and exchange interactions, whose strengths can be amplified by tuning the parametric gain. Meanwhile, the enhanced interaction strengths can be smoothly tuned by various physical parameters, such as the separation between giant atoms and the phase difference of counter-propagating pump fields.
	
	Moreover, our proposal intrinsically eliminates next-nearest-neighbor couplings, thereby enabling accurate simulation of many-body quantum phase transitions. This architecture also demonstrates the potential of giant atoms in mitigating squeezing-induced decoherence. This work provides a promising approach toward realizing strongly interacting many-body physics with inherent coherence protection and engineering quantum dynamics in waveguide QED systems. 
	
	\section{Acknowledgments} 
	We thank Dr.~Tao Liu and Dr. Zi-Yong Ge for insightful discussions and valuable suggestions on this work. X. W. is supported by the National Natural Science Foundation of China (NSFC) (No.~1217430), and the Fundamental Research Funds for the Central Universities (No. xzy012023053).
	
	\appendix
	\section{ SLH Formalism for Giant Atoms}
	\label{appendix_a}
	\begin{figure*}
		\centering
		\includegraphics[width=0.95\linewidth]{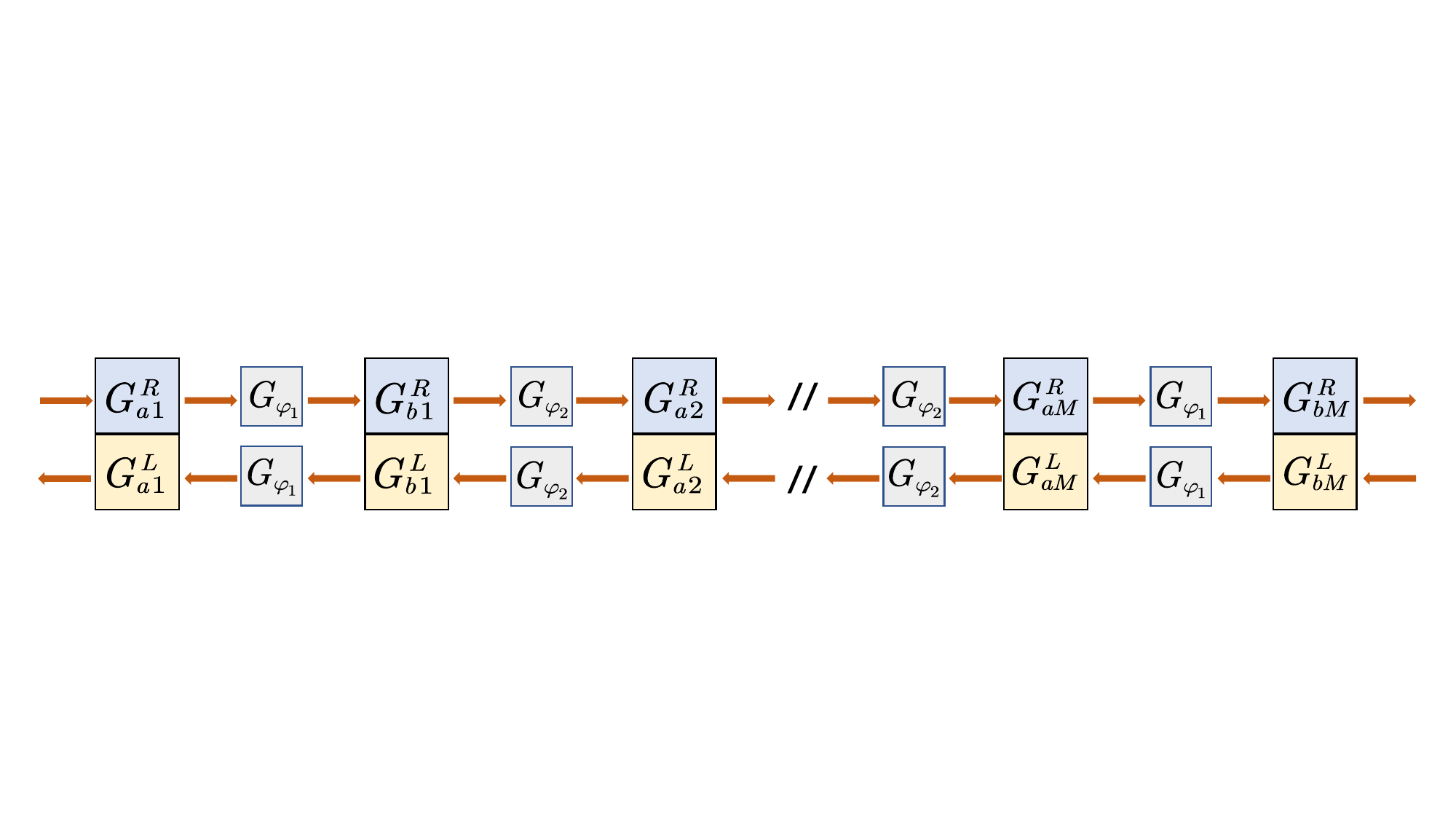}
		\caption{The input-output flows in the corresponding SLH calculation.}
		\label{fig8}
	\end{figure*}
	We present the derivation process of SLH formalism for a two braided giant atoms system with $M$ coupling points in Fig.~\ref{fig8}. The triplets for the right- and left-moving component are defined as
	\begin{widetext}
		\begin{align}
			G^R=(\mathcal{S}_R,\mathcal{L} _R,H_R)=G_{bM}^{R}\lhd G_{\varphi _1}\lhd G_{aM}^{R}\lhd G_{\varphi _2} \lhd G_{b2}^{R}\lhd G_{\varphi _1}\lhd G_{a2}^{R}\lhd  \cdots \lhd  G_{\varphi _2}\lhd G_{b1}^{R}\lhd G_{\varphi _1}\lhd G_{a1}^{R}, \notag\\
			G^L=(\mathcal{S}_L,\mathcal{L} _L,H_L)=G_{a1}^{L}\lhd G_{\varphi _1}\lhd G_{b1}^{L} \lhd G_{\varphi _2}\lhd G_{a2}^{L}  \lhd G_{\varphi _1}\lhd G_{b2,L}\lhd \cdots \lhd G_{\varphi _2}\lhd G_{aM}^{L}\lhd G_{\varphi _1}\lhd G_{bM}^{L}.
		\end{align}
	\end{widetext}
	where $\varphi_1=k_0(z_{b_i}-z_{a_i})$ and $\varphi_2=k_0(z_{a_{i+1}}-z_{b_i})$ represent the phase difference between the adjacent coupling points, with $\varphi_1+\varphi_2=\pi$. The atomic triplets for right-propagating modes are divided into the following components
	\begin{eqnarray}
		&&G_{\varphi_{1(2)}} = \left( e^{i\varphi_{1(2)}}, 0, 0 \right),\notag 
		\\	&&G_{nM}^R = \left( 1,\sqrt{\frac{\gamma_{nM}}{2}} S_R(z_{nM}), 0 \right), \notag \\
		&&G_{n1}^R = \left( 1,\sqrt{\frac{\gamma_{n1}}{2}} S_R(z_{n1}),  \frac{\Delta_n}{2}\sigma_z^n \right).
	\end{eqnarray}
	By repeatedly applying the series product rule, we obtain $G^R$ as
	\begin{align}
		\mathcal{S}_R &= e^{i[M\varphi_1 + (M-1)\varphi_2]},  \notag \\
		\mathcal{L}_R &= e^{i[ M\varphi_1 + (M-1)\varphi_2 ]} \sqrt{\frac{\gamma_{a1}}{2}}\, S_R(z_{a1}) \notag \\
		& + e^{i[ (M-1)\varphi_1 + (M-1)\varphi_2 ]} \sqrt{\frac{\gamma_{b1}}{2}}\, S_R(z_{b1}) + \cdots \notag  \\
		& + e^{i\varphi_1} \sqrt{\frac{\gamma_{aM}}{2}}\, S_R(z_{aM})
		+ \sqrt{\frac{\gamma_{bM}}{2}}\, S_R(z_{bM}),
	\end{align}
	with Hamiltonian $H_R=\frac{\Delta_a}{2}\sigma_z^{a} + \frac{\Delta_b}{2}\sigma_z^{b}+H_{\text{int,R}}$:
	\begin{align}
		H_{\text{int,R}} &=\sum_{p,q} \frac{\sqrt{\gamma_{ap}\gamma_{bq}}}{4i}
		\operatorname{sgn}\bigl( z_{bq} - z_{ap} \bigr)  \notag \\
		&\times \Big[ e^{i( \phi_{bq}-\phi_{ap} )} S_R^{\dagger}( z_{bq} ) S_R( z_{ap} ) \notag\\
		&- e^{-i( \phi_{bq}-\phi_{ap} )} S_R^{\dagger}( z_{ap} ) S_R( z_{bq} ) \Big].
		\label{H_intR}
	\end{align}
	By substituting $S_R(z)$ into Eq.~(\ref{H_intR}) and using the sum-to-product identities, we obtain
	\begin{gather}
		H_{R} =\frac{\Delta_a}{2}\sigma_z^{a}\! +\! \frac{\Delta_b}{2}\sigma_z^{b}\!+\!\frac{\sqrt{\gamma_{ap}\gamma_{bq}}}{4i}\sum_{p,q} \mathrm{sgn}\!\left( z_{bq} - z_{ap} \right)\notag \\ \times\Big[
		e^{i\left( \phi_{bq} - \phi_{ap} \right)} \cosh\!\left[ \tfrac{\mathcal{G}}{2}\left( z_{bq} - z_{ap} \right) \right] \sigma_{+}^{b}\sigma_{-}^{a} \notag\\
		- e^{i\left( \phi_{bq} + \phi_{ap} \right)} e^{i\vec{\theta}} \sinh\!\left[ \tfrac{\mathcal{G}}{2}\left( z_{bq} - z_{ap} \right) \right] \sigma_{+}^{a}\sigma_{+}^{b}+\mathrm{H}.\mathrm{c}. \Big].
	\end{gather}
	
	The atomic triplets for left-propagating modes are divided into the following components
	\begin{align}
		G_{n1}^L &= \left( 1, \sqrt{\frac{\gamma_{n1}}{2}} S_L(z_{n1}), 0 \right), \notag \\
		G_{nM}^L &= \left( 1, \sqrt{\frac{\gamma_{nM}}{2}} S_L(z_{nM}), 0 \right).
	\end{align}
	Based on the evident symmetry shown in Fig.~\ref{fig8}, it can be directly inferred that the triplet associated with the left-moving component, $G_L=(\mathcal{S}_L,\mathcal{L}_L,H_L)$:
	\begin{align}
		\mathcal{S}_L &= e^{i\left[ M\varphi_1 + (M-1)\varphi_2 \right]},  \notag\\
		\mathcal{L}_L &= e^{i\left[ M\varphi_1 + (M-1)\varphi_2 \right]} \sqrt{\frac{\gamma_{bM}}{2}}\,S_L(z_{bM}) \notag\\
		& + e^{i\left[ (M-1)\varphi_1 + (M-1)\varphi_2 \right]} \sqrt{\frac{\gamma_{aM}}{2}}\,S_L(z_{aM}) \notag\\
		& + \cdots 
		+ e^{i\varphi_1} \sqrt{\frac{\gamma_{b1}}{2}}\,S_L(z_{b1})
		+ \sqrt{\frac{\gamma_{a1}}{2}}\,S_L(z_{a1}).
	\end{align}
	and the Hamiltonian reads
	\begin{gather}
		H_{L} =\frac{\sqrt{\gamma_{ap}\gamma_{bq}}}{4i}\sum_{p,q} \mathrm{sgn}\!\left( z_{bq}\! -\! z_{ap} \right)\times\notag \\ \Big[ -e^{-i\left( \phi_{bq} + \phi_{ap} \right)} e^{i\vec{\theta}} \sinh\!\left[ \tfrac{\mathcal{G}}{2}\left( z_{bq} - z_{ap} \right) \right] \sigma_{+}^{b}\sigma_{+}^{a}  \notag\\
		\quad - e^{-i\left( \phi_{bq} - \phi_{ap} \right)} \cosh\!\left[ \tfrac{\mathcal{G}}{2}\left( z_{bq} - z_{ap} \right) \right] \sigma_{+}^{b}\sigma_{-}^{a} +\mathrm{H}.\mathrm{c}.\Big]
	\end{gather}
	
	To eliminate the decoherence interactions in the system, we require $\mathcal{L}_R=\mathcal{L}_L=0$, and obtain
	\begin{gather}
		\begin{cases}
			\sum_{i=1}^{M}{\sqrt{g\left( z_{ni} \right)}e^{i\left( M-i \right) \pi}\cosh \left( \frac{\mathcal{G}}{2}z_{ni} \right)}=0,\\
			\sum_{i=1}^{M}{\sqrt{g\left( z_{ni} \right)}e^{i\left( M-i \right) \pi}\sinh \left( \frac{\mathcal{G}}{2}z_{ni} \right)}=0.\\
		\end{cases}
	\end{gather}

%

\end{document}